\documentstyle[12pt,a4]{article}

\def\beq{\begin{equation}}
\def\eeq{\end{equation}}

\begin{document}

\begin{flushright}

Crete/02-12 \\
ITEP/TH-33/02 \\

\end{flushright}

\vspace{0.5cm}

\begin{center}

{\Large\bf
Brane-induced decay of the Kaluza-Klein vacuum}\\

\end{center}

\bigskip

\begin{center}

{\bf K. Selivanov$^{1}$ and T.N. Tomaras$^{2,3}$}

\bigskip

{\sl $^{1}$ ITEP, B. Cheryomushkinskaya 25, 117218 Moscow, Russia}

{\sl $^{2}$ Department of Physics and Institute of Plasma Physics, University of Crete,}

{\sl 71003 Heraklion, Greece}

{\sl $^{3}$ Foundation of Research and Technology Hellas, 71110 Heraklion, Greece}

{E-mail: selivano@heron.itep.ru, tomaras@physics.uoc.gr}

\end{center}

\bigskip

\begin{abstract}
The enhancement in the decay rate of the Kaluza-Klein vacuum due to the presence of
a brane is studied, both in the test brane approximation and beyond it. 
Spontaneous brane materialization in the Kaluza-Klein vacuum
is also described. 
\end{abstract}

\section{Introduction}
It was shown long time ago that the Kaluza-Klein
theory is non-perturbatively unstable \cite{Witten}. In the simplest
case, $R^4 \times S^1$ space can decay via the formation of the, so called,
``bubbles of nothing''. The bounce
solution describing the decay was shown to be the euclidean 5d black hole.
Later on, this process was generalized to the Melvin
type Kaluza-Klein vacuum with magnetic flux \cite{Dowker1}. The vacuum
decays via the mediation of the euclidean 5d Kerr solution,
which is the corresponding bounce configuration in this case.

Such manifold decay processes have also been discussed in the context
of string/M theory \cite{horava, costa}, while the whole subject has become 
quite popular recently, partially because
of the interest in string dynamics in time-dependent 
backgrounds \cite{Aharony - Bala}.

The aim of the present paper is to study the induced decay of
Kaluza-Klein vacuum, that is, the decay caused by the presence of a
heavy object (particle, cosmic string, 2-brane) in the initial state.
It is natural to expect that the presence of a heavy object in the
initial state enhances the decay rate. A qualitative explanation of
this effect for the case of particle induced decay was given in
\cite{Selivanov1}: the worldline of the inducing particle
in some circumstances happens to be ``shorter'' in the presence
of the bounce than in its absence and this leads to the enhancement factor
$e^{2Rm}$ for the decay rate where $R$ is the radius of the bounce.
In fact, as explained in \cite{Selivanov1}, for a sufficiently
heavy particle one should take into account the back reaction of the particle
onto the bounce. The resulting distortion of the latter gives rise to an 
even stronger enhancement. These ideas were further developed to 
describe the nonperturbative instability of neutral
branes in external fields \cite{Gorsky1}, or 
the spontaneous creation of the so-called Brane-World \cite{Gorsky2}.

The present paper is organized in six sections, of which 
this introduction is the first.
In Section 2 we briefly review the case of spontaneous decay of the Kaluza-Klein
$R^4\times S^1$ vacuum.

In section 3 we compute the enhancement of this decay rate, due to the
presence of a 2-brane (codimension 1), a 1-brane (codimension 2) or 
a 0-brane (codimension 3) in the, so-called,
test brane approximation.
In this approximation the back reaction of the external object onto the 
background geometry
is not taken into account and it is easy to treat. Essentially, one 
has to compare areas of worldsheets
of the inducing brane in the presence and the absence of bounce.
Inducing branes of arbitrary codimension are discussed.
Actually, codimension 1 case happens to be degenerate - the brane gives an ill-defined
boundary dependent contribution to the decay rate. We attribute this to the fact
that codimension 1 brane beyond the test brane approximation does not
exist in the asymptotically locally flat space. The cases of codimension 2 and
3 are unambiguous.

In section 4 we go beyond the test brane approximation, but only 
for the case of the codimension 2 brane. 
A ``bubble of nothing'' is formed, which breaks the brane and expands
to fill out the whole space. 
The bounce configuration relevant to this case and the computation of the
enhancement factor of the vacuum decay rate are presented here. It so happens 
that the enhancement factor coincides with the one obtained in the 
test brane approximation. 
The process discussed here is different from those considered in 
\cite{Dowker}--\cite{Eardley}, where the external string breaks and two
black holes are formed.

Somewhat outside the main theme of this paper, in section 5 we consider 
not induced but spontaneous decay of the Kaluza-Klein vacuum 
via a channel with a 2-brane in the final state.
A minor modification of Witten's construction allows
to describe spontaneous decay of the vacuum
into a bubble of nothing with a codimension 2 brane on its boundary.

Section 6 contains our conclusions.

\section{Spontaneous decay of $R^4\times S^1$.}

Let us consider the locally flat manifold $R^4 \times S^1$
with metric
\beq
\label{flat}
ds^2=dr^2+r^2d\Omega_3^2+d\tau^2
\eeq
where $d\Omega_3^2$ is the metric on $S_3$ and $\tau$ runs over
the circle,
\beq
\label{circle}
\tau \sim \tau+2\pi R.
\eeq
This manifold was shown to decay \cite{Witten}, the corresponding bounce
being described by the metric
\beq
\label{bubble}
ds^2=\Bigl(1-\frac{\alpha}{r^2}\Bigr)^{-1} dr^2+r^2d\Omega_3^2+
({1-\frac{\alpha}{r^2}})d\tau^2.
\eeq
This metric is nothing but the 5d Schwarschild metric
analytically continued to euclidean space. The coordinate $\tau$
runs over the same circle (\ref{circle}), while
$r$ runs from $\sqrt{\alpha}$ to $\infty$. This does not mean, 
that $r=\sqrt{\alpha}$ is a boundary of the manifold (\ref{bubble}).
In fact, $r=\sqrt{\alpha}$  defines a codimension 2 subspace (because $S^1$ spanned by
$\tau$ shrinks to a point on this subspace). Geometrically it is an $S_3$.
Generically, this $S_3$ is a locus of conical
singularity unless $\alpha$ and $R$ are related by
\beq
\label{relation}
\alpha=R^2.
\eeq
The surface
\beq
\label{hole}
r=\sqrt{\alpha}
\eeq
as well as the whole geometry described by (\ref{bubble}) is sometimes
called ``bubble of nothing''. It is interpreted as a bounce, since it has one 
negative mode in the spectrum of small fluctuations around it \cite{Witten}. 

As usual in the semiclassical approximation, the decay rate is with exponential
accuracy estimated as $\exp{(-\Delta S)}$ where
$\Delta S\equiv S-S_0$, $S$ being the euclidean action
of the relevant bounce and $S_0$ the action of the reference background 
$R^4 \times S^1$ with metric (\ref{flat}), 
the natural reference space for the discussion of spontaneous 
decay of the Kaluza-Klein vacuum.
The euclidean action is
\beq
\label{action}
S=-\frac{1}{16\pi G_5} \int \sqrt{g}R -
\frac{1}{8\pi G_5} \oint \sqrt{h} K,
\eeq
where the first term is the Einstein-Hilbert term,
and the second is the integral over the boundary of the trace $K$ of the extrinsic
curvature of the boundary surface, with
the measure constructed from the induced metric $h$.
In non-compact spaces this action is infinite. However, one needs only
the difference $\Delta S\equiv S-S_0$ which, being properly defined, is finite.
An important point here is that in the computation of
$S_0$ one has to consider a boundary with the same intrinsic geometry as in the
computation of $S$. Given the metric (\ref{bubble}), it is convenient to take
the boundary 
\beq
\label{boundary}
r=r_0,
\eeq
where $r_0$ is large compared to all physical length scales.
Restricting metric (\ref{bubble}) to the surface (\ref{boundary})
gives the induced metric $h$ on the boundary,
\beq
\label{bmetric}
h_{ij}dx^idx^j=r_0^2d\Omega_3^2+
({1-\frac{\alpha}{r_0^2}})d\tau^2.
\eeq
Then, to find in the reference space a boundary surface with the same intrinsic
geometry (\ref{bmetric}),
one should take in $R^4 \times S^1$ the same surface (\ref{boundary})
and - what is more subtle - tune the period of $\tau$, so that
\beq
\label{circle1}
\tau \sim \tau+2\pi R \sqrt{1-\frac{\alpha}{r_0^2}}.
\eeq

The rest of the calculation is straightforward. The curvature term does not contribute
because both metrics (\ref{bubble}) and (\ref{flat}) are Ricci flat.
The trace of the extrinsic curvature can be computed using the identity
\beq
\label{ext}
K\sqrt{h}=n\sqrt{h},
\eeq
where $h$ is the determinant of the induced metric (\ref{bmetric}) on the boundary,
\beq
\label{h}
\sqrt{h}=r_0^3 \sqrt{1-\frac{\alpha}{r_0^2}}\sin^2\theta_1|\sin\theta_2|
\eeq
(on $S_3$ we use the standard spherical coordinates (\ref{spherical}) below), 
and $n$ is the unit normal
\beq
\label{normal}
n=\sqrt{1-\frac{\alpha}{r_0^2}} \frac{\partial}{\partial r_0}.
\eeq
This way we obtain the boundary term in the ``bubble'' space
\beq
K\sqrt{h}=3r_0^3(1-\frac{2}{3}\frac{\alpha}{r_0^2})
\sin^2\theta_1|\sin\theta_2|.
\eeq
The simplest way to evaluate the trace of the extrinsic curvature $K_0$ of
the boundary (\ref{boundary})
in the reference space $R^4\times S^1$, is to note that it is
equal to the sum of principal curvatures of $S_3$ of radius $r_0$
i.e. $K_0=3/r_0$. The measure $\sqrt{h}$ is of course the same
on the boundary in the reference space and is given by (\ref{h}), since
the boundaries were tuned to have the same intrinsic geometry. 

Finally, the difference of the boundary integrals gives
\beq
\label{action1}
\Delta S=-\frac{1}{8\pi G_5}\int (K-K_0)\sqrt{h}=\frac{\pi R^2}{8G_4},
\eeq
where $G_4=2\pi R G_5$.

We conclude that in the exponential approximation 
the probability per unit time and per unit volume of formation of a 
critical bubble is equal to
\beq
\label{rate}
\Gamma_0 / V \sim G_4^{-2} e^{-\frac{\pi R^2}{8G_4}}.
\eeq
In fact, the evaluation of the prefactor requires the computation of a 
functional determinant, usually a formidable task \cite{Kiselev}.
In the problem at hand, however,
the prefactor is $G_4^{-2}$ times a function of the dimensionless ratio 
$R^2/G_4$. We arbitrarily, but consistently with the exponential
approximation discussed here, took this function equal to one. 
The probability of formation of such a critical bubble in the volume 
of the observable Universe and within its age is, as it should be, 
much smaller than unity,
as long as the size $R$ of the extra dimension is 
greater than or about $40 M_{Pl}^{-1}$. Thus, the spontaneous decay of
the Kaluza-Klein vacuum does not imply a severe constraint on the size
of the internal dimension.

\section{Induced decay in the test brane approximation}

We now turn to the computation
of the enhancement of the vacuum decay rate due to the presence of 
a test brane. We shall treat the codimension 1, 2 and 3 cases separately.
In the next section we shall deal with the full problem,
including back reaction
on the background metric due to the brane.  
The dynamics of the brane is described by the Nambu-Goto action, which for
a p-brane has the form
\beq
S_{NG}=T_p\int d^{p+1}\xi \sqrt{\det g_{ind}},
\label{nambu}
\eeq
where $T_p$ is the brane tension, a positive constant with dimensions of (mass)$^{p+1}$.
A natural assumption, implicit in the discussion of test branes in the Kaluza-Klein vacuum,
is that the presence of the branes does not remove the negative mode from the 
spectrum of small fluctuations around the bounce configuration, responsible for
the instability of the pure vacuum. Their only effect in the exponential approximation 
is to modify the decay rate by a term in the exponent 
linear in the brane tension.

\subsection{Codimension 1 test brane.}

The application of the test brane
approximation is questionable in this case, because a codimension 1 brane 
is not compatible with asymptotic flatness \cite{Vilenkin}. Nevertheless, 
we would like to discuss it just to demonstrate in a simple example the method we
shall be using and to see what exactly will go wrong with our approximation.

Let us consider the same $R^4 \times S^1$ space (\ref{flat})
and a 3-brane with tension $T_3$ located, say, at $x=0$,
where $x$ is one of the cartesian coordinates in $R^4$, and
wrapped around $S^1$. In the context of the test brane approximation
we ignore the backreaction of the brane on the geometry of the background,
which is assumed to be the flat space (\ref{flat}). The brane dynamics is
described by the Nambu-Goto action (\ref{nambu}) with $p=3$ and 
the enhancement factor to leading order 
is the exponential of the difference of the actions
of the brane in the bounce (\ref{bubble}) and in the flat geometry
(\ref{flat}). 
Thus, to describe the influence of the brane on the vacuum decay rate, 
we have to find the minimal surface in the bounce
geometry (\ref{bubble}), which coincides asymptotically with the worldsheet
of the brane in flat space. This is easiest in the 
spherical coordinates $0\leq \theta_1, \theta_2 \leq \pi$ 
and $0\leq \theta_3\leq 2\pi$, 
in terms of which the $d{\Omega}_3^2$ term in (\ref{flat}) and (\ref{bubble}) 
takes the standard form
\beq
\label{spherical}
d{\Omega}_3^2=d\theta_1^2+ \sin^2 \theta_1 d\theta_2^2 +
\sin^2 \theta_1 \sin^2 \theta_2 d\theta_3^2.
\eeq
Consider the surfaces
\beq
\label{c1}
\theta_1=\pi/2
\eeq
in the spaces (\ref{flat}) and (\ref{bubble}).
By symmetry these are minimal surfaces. These surfaces coincide asymptotically
$r\to\infty$ and are asymptotically of the required type. Thus (\ref{c1})
describes the worldsheet of the inducing brane.

Let us clarify the geometry of the brane near the bubble.
The brane intersects the 3-sphere (\ref{hole}) along an equatorial $S_2$.
Since the $S^1$ spanned by $\tau$ shrinks to a point on this sphere,
we conclude that there is a bubble of nothing inside the brane too.

It remains only to compute the difference of the Nambu-Goto actions
of the brane in the presence and in the absence of the bubble.
Note that the worldsheet volumes are infinite in both cases, so we have to use
a boundary as it was discussed in the previous section.
Notice that the $1-\alpha/r^2$ factors cancel out in the determinant,
so that the difference in the actions is simply
\beq
\label{actionc1}
\Delta S_{NG}=4 \pi T_3 2\pi R
\left( \int_{\sqrt{\alpha}}^{r_0}r^2dr -
\sqrt{1-\frac{\alpha}{r_0^2}}\int_{0}^{r_0}r^2dr \right) 
\sim {4\over 3}\pi^2 T_3 R\alpha r_0.
\eeq
The factor in front of the second integral is due to the period
rescalling (\ref{circle1}) of $\tau$ in flat space. 
Notice, that $\Delta S_{NG}$ in (\ref{actionc1}) diverges
linearly when $r_0 \rightarrow \infty $. 
One would be tempted to conclude at this point that the presence of the 
brane stabilizes the flat Kaluza-Klein space-time. However,
$\Delta S_{NG}$ obtained above is ill-defined, being boundary dependent and divergent.
This behaviour is a manifestation of the fact mentioned in the beginning that
the presence of a codimension 1 brane affects drastically the asymptotic 
behaviour of the background, which changes into AdS.

\subsection{Codimension 2 test brane.}

Consideration of this case is quite similar to the previous one
apart from some technical details. The inducing brane is of
codimension 2 in the $R^4$ component and is wrapped around the
$S^1$ spanned by $\tau$.
In this case, a convenient coordinate system to use on $S_3$ is 
\beq
\label{spherical2}
d{\Omega}_3^2= \sin^2 \theta d\psi^2 + d\theta^2 + \cos^2 \theta d\phi^2,
\eeq
where $\theta$ runs from $0$ to $\pi/2$, while $\psi$
and $\phi$ run from $0$ to $2\pi$.
Consider the surfaces
\beq
\label{c2}
\theta=0
\eeq
in spaces (\ref{flat}) and (\ref{bubble}).
The $S^1$ spanned by $\psi$ shrinks to a point on these surfaces, which
consequently have codimension 2. 
By symmetry they are minimal surfaces. 
In addition, these surfaces again coincide asymptotically
and are asymptotically of the required type. Thus (\ref{c2})
describes the worldsheet of the inducing brane.
Similarly to the codimension 1 case above, in the presence of the bubble of nothing
in the space, there is a ``hole" of nothing on the inducing brane too.
After its formation the critical bubble and with it the hole on the brane 
expands, eliminating the brane and pushing
space to infinity. 

The computation of the enhancement factor is the same as in the codimension 1
case. The difference of the Nambu-Goto actions is
\beq
\label{actionc2}
\Delta S_{NG}=2\pi T_2 2\pi R
\left(\int_{\sqrt{\alpha}}^{r_0}rdr -
\sqrt{1-\frac{\alpha}{r_0^2}}\int_{0}^{r_0}rdr \right)
\sim -\pi^2 \alpha^{3/2} T_2 \, ,
\eeq
so that the enhancement factor is equal to
\beq
\label{enc2}
\Gamma/{\Gamma}_0=e^{\pi^2 \alpha^{3/2} T_2},
\eeq
where $T_2$ is the tension of the inducing brane. 
In terms of $R$ satisfying (\ref{relation}) and the 
``macroscopic" tension $\mu_2\equiv 2\pi R T_2$ of the brane, i.e. the 
energy of the brane per unit of non-compact volume, 
the enhancement factor takes the form:
\beq
\label{enc21}
\Gamma/{\Gamma}_0=e^{\frac{1}{2} \pi R^2 \mu_2}.
\eeq

\subsection{Codimension 3 test brane.}

In the case of codimension 3 the decay is induced by a 1-brane, a string
wrapped around the compact dimension.
In this case instead of (\ref{actionc2}) we obtain
\beq
\label{actionc3}
\Delta S=2 T_1 2\pi R
\left( \int_{\sqrt{\alpha}}^{r_0}dr -
\sqrt{1-\frac{\alpha}{r_0^2}}\int_{0}^{r_0}dr \right).
\eeq
The factor $\sqrt{1-\frac{\alpha}{r_0^2}}$ does not contribute in this case,
and we end-up with
\beq
\label{enc3}
\Gamma/{\Gamma}_0=e^{4 \pi \alpha T_1}.
\eeq
Defining $\mu_1\equiv 2 \pi R T_1$ and using equation (\ref{relation})
one arrives at
\beq
\label{enc31}
\Gamma/{\Gamma}_0=e^{2R\mu_1}.
\eeq

\section{Beyond the test brane approximation}

To go beyond the test brane approximation of the previous section,
one needs to find
a solution of the Einstein-Nambu-Goto equations, which is a deformation
of the bubble of nothing metric (\ref{bubble}) due to the presence
of the brane. We shall consider here only the case of codimension 2
brane. Other cases will be considered elsewhere.

The metric describing a codimension 2 brane in $R^4 \times S^1$
is characterized by the so-called deficit angle:
\beq
\label{string}
ds^2=dr^2+
r^2 (\nu^2 \sin^2 \theta  d\psi^2 + d\theta^2 + \cos^2 \theta d\phi^2)
+d\tau^2,
\eeq
where, as discussed above, $\theta$ runs from $0$ to $\pi/2$, $\psi$
and $\phi$ run from $0$ to $2\pi$ and $\nu$ is a constant
less than 1.
The metric in equation (\ref{string}) is singular at $\theta=0$.
This is where the brane is located.
In the orthogonal
plane, $\theta=\pi/2$, there is a deficit angle equal to
$\Delta=2\pi(1-\nu)$.
It is well known \cite{Vilenkin}
that the deficit angle is related to the brane tension $T_2$ according to
\beq
\label{deficit}
\nu=1-4 G_5 T_2 = 1-4 G_4 \mu_2.
\eeq
Strictly speaking, only the string case, that is, codimension 2
brane in 3+1-dimensions, is considered in \cite{Vilenkin}, but it is only
codimension that matters.

Consider the metric
\beq
\label{string2}
ds^2=(1-\frac{\alpha}{r^2})^{-1} dr^2+
r^2(\nu^2 \sin^2 \theta  d\psi^2 + d\theta^2 + \cos^2 \theta d\phi^2)
+({1-\frac{\alpha}{r^2}})d\tau^2.
\eeq
It is a solution of the Einstein-Nambu-Goto field equations, and represents a 
generalization of (\ref{string}) describing both the bubble and
the string and which coincides with (\ref{string}) asymptotically as $r\to\infty$.
It is, of course, singular along the locus of the brane, and
since we do not want any other singularity, we require
the parameter $\alpha$ to be related to the Kaluza-Klein radius by 
(\ref{relation}).

The geometry of the brane near the bubble is
the same as in the test brane case: there is a bubble of nothing inside
the brane which is an intersection of the brane with the bubble of nothing in
the total space.

It remains to estimate the induced decay rate which with
exponential accuracy is equal to
the exponential of $-\Delta S$, the difference of actions of the metrics
(\ref{string2}) and (\ref{string}).
The action (\ref{action}) is now modified to include the Nambu-Goto
term
\beq
\label{action2}
S=-\frac{1}{16\pi G_5} \int \sqrt{g}R + T_2 \oint \sqrt{g_{ind}}-
\frac{1}{8\pi G_5} \oint \sqrt{h} K.
\eeq
Note that because of the conical singularity the $R$-term is not
zero, rather it gives a contribution similar to the Nambu-Goto term.
We now would like to argue that the first two terms cancel each other
in the case of the codimension 2 brane.
One way to show this is to smoothen-out the brane in the transverse directions, to solve
Einstein equations and to see explicitly that the contributions
cancel. This was performed in \cite{Vilenkin} for the case of a straight string.
Contributions due to non-straightness of the string should arise only
for a string with finite thickness. Actually, cancellation of the first two terms
is equivalent to nothing but the relation (\ref{deficit}).

Thus we are  left with the difference of the boundary terms. The computation
then follows the lines reviewed in section 2. 
The parameter $\nu$ enters only in $h$; $K$ is $\nu$-independent.
Thus, the final answer for $\Delta S$ differs from (\ref{action1}) just
by a factor of $\nu$:
\beq
\label{action3}
\Delta S=\nu\frac{\pi R^2}{8G_4}.
\eeq
We conclude that the probability per unit of space-time volume $V$ of the 
formation of a critical bubble in the Kaluza-Klein state in the presence of 
the codimension 2 brane is equal to
\beq
\Gamma/V \sim e^{-\nu\frac{\pi R^2}{8G_4}}.
\eeq
Notice that upon substitution of $\nu$ in terms of the brane tension from (\ref{deficit})
we recover the result (\ref{enc21}), obtained in the test brane approximation.

\section{Spontaneous vacuum decay with a brane in the final state.}

As a byproduct of the previous discussion, in this section we shall
describe a new channel of spontaneous decay 
of the Kaluza-Klein vacuum. It is characterized by the existence of 
a brane in the final state. 
Consideration of this subsection is inspired by the previous section,
where the metric with a conical singularity was used to describe
the inducing codimension 2 brane. Now recall that the metric (\ref{bubble})
has a conical singularity at $r=\sqrt{\alpha}$ when the parameter
$\alpha$ is not tuned according to equation (\ref{relation}).
Indeed, changing in (\ref{bubble}) the coordinates so that
$r\equiv \sqrt{\alpha}+{\rho^2}/{2 \sqrt{\alpha}}$ we arrive in the vicinity
of $\rho=0$ to the metric
\beq
\label{conical}
ds^2=d\rho^2+\frac{\rho^2}{\alpha}d\tau^2 + \alpha d\Omega_3^2.
\eeq
Thus, in view of (\ref{circle}), the deficit angle 
parameter $\nu$ is given by 
\beq
\nu=\frac{R}{\sqrt{\alpha}}.
\eeq
The presence of the conical singularity represents a codimension 2 brane, and
equation (\ref{deficit}) relates the parameter $\alpha$, the Kaluza-Klein radius $R$
and the tension of this brane $T_2$ by
\beq
\label{relation2}
\frac{R}{\sqrt{\alpha}}=1-4G_5 T_2.
\eeq
The brane is located at $r=\sqrt{\alpha}$ and its euclidean signature 
world-sheet is $S^3$ with
induced metric 
\beq
ds_{ind}^2=\alpha d\Omega_3^2.
\label{induced}
\eeq 
According to \cite{Witten}, the 
critical bubble is the $\theta=\pi/2$ section of the bounce. Thus, at ``time'' 
$\theta=\pi/2$ a brane with the geometry of $S^2$ materializes in space 
and starts expanding. 

So, if in the ``spectrum" of the theory under consideration there exist
codimension 2 branes
with tension $T_2$, the Kaluza-Klein vacuum can decay
not only via the bounce (\ref{bubble}) with $\alpha$ obeying
(\ref{relation}), but also via bounce (\ref{bubble})
with $\alpha$ obeying (\ref{relation2}). This is an additional channel
of spontaneous decay. 
We should point out here, that there is no problem with energy conservation.
The discussion of \cite{Witten} applies to our case as well. 
Both the initial and the final states have zero energy, since asymptotically
the departure of the metric from flat space does not contain $1/r$ terms.
 
To estimate the rate of spontaneous decay in this channel, we follow 
the steps reviewed in
section 2. In fact, one only has to distinguish between $\alpha$ and $R^2$.
The result is
\beq
\label{result}
\Gamma/V \sim e^{-\frac{\pi R^2}{8G_4(1-8\pi R G_4 T_2)^2}}.
\eeq

Note that this result is compatible with the test brane approximation of the
same process. Indeed, in this approximation the correction in the exponent of
$\Gamma$ would be the Nambu-Goto action of the test brane with tension $T_2$
located at $r=\sqrt{\alpha}$ in the manifold (\ref{bubble}). 
The Nambu-Goto action of the induced metric
on the world-sheet of the brane (\ref{induced}) is $2\pi^2 R^3 T_2$. 
This is identical to the linear in $G_5 T_2$ term
in the expansion of the exponent of (\ref{result}).

\section{Conclusion}

We studied how the Kaluza-Klein vacuum - that is the space-time of the type
$R^4\times S^1$ - decays in the presence of branes. As it is natural to expect,
the presence of branes generically facilitates the decay. 
The enhancement factor was computed,
to exponential accuracy, first in the test brane approximation
(when one neglects back reaction of the brane onto the gravity), as well as beyond
the test brane approximation in the case of codimension 2 brane only.
We also considered the possibility of having a 
codimension 2 brane (membrane)
in the final state of the spontaneous decay of the Kaluza-Klein vacuum and
estimated the rate of decay in this channel.

\vspace{1.5cm}

{\small
\noindent
{\normalsize\bf Acknowledgements}

\medskip
\noindent
We would like to thank A. Gorsky for collaboration at an early stage of this work,
for useful discussions and for reading the manuscript.
K.S. acknowledges the Physics Department of the University of Crete,
where this work was done, for its kind hospitality.
This work was partially supported by European Union grants HPRN-CT-2000-00122 and 
-00131. 
K.S. was supported in part by INTAS-00-334, by Grant RFBR 00-15-96562 
for the support of scientific schools, as well
as by a NATO travel grant. 
}


\begin{thebibliography}{100}

\bibitem{Witten}
E.~Witten,
Nucl.\ Phys.\ B {\bf 195}, 481 (1982).

\bibitem{Dowker1}
F.~Dowker, J.~P.~Gauntlett, G.~W.~Gibbons and G.~T.~Horowitz,
Phys.\ Rev.\ D {\bf 52}, 6929 (1995)
[arXiv:hep-th/9507143].


\bibitem{horava}
M. Fabinger and P. Horava,
Nucl. Phys. {\bf B 580}, 243 (2000)

\bibitem{costa}
J. Russo and A. Tseytlin,
Nucl. Phys. {\bf B461}, 131, (1996)  \\
M.~S.~Costa and M.~Gutperle,
JHEP {\bf 0103}, 027 (2001)
[arXiv:hep-th/0012072] \\
M. Gutperle and A.Strominger,
JHEP {\bf 0106}, 035 (2001) \\
A. Tseytlin,
[arXiv:hep-th/0108140] \\
S. De Alwis and A. Flournoy,
[arXiv:hep-th/0201185]



\bibitem{Aharony}
O.~Aharony, M.~Fabinger, G.~T.~Horowitz and E.~Silverstein,
arXiv:hep-th/0204158.



\bibitem{Birmin}
D.~Birmingham and M.~Rinaldi,
arXiv:hep-th/0205246.

\bibitem{Bala}
V.~Balasubramanian and S.~F.~Ross,
arXiv:hep-th/0205290.


\bibitem{Selivanov1}
K.~G.~Selivanov and M.~B.~Voloshin,
JETP Lett.\  {\bf 42}, 422 (1985).

\bibitem{Gorsky1}
A.~Gorsky and K.~Selivanov,
Nucl.\ Phys.\ B {\bf 571}, 120 (2000)
[arXiv:hep-th/9904041].

\bibitem{Gorsky2}
A.~Gorsky and K.~Selivanov,
Phys.\ Lett.\ B {\bf 485}, 271 (2000)
[arXiv:hep-th/0005066].\\
A.~S.~Gorsky and K.~G.~Selivanov,
Int.\ J.\ Mod.\ Phys.\  {\bf 16}, 2243 (2001)
[arXiv:hep-th/0006044].


\bibitem{Dowker}
F.~Dowker, J.~P.~Gauntlett, G.~W.~Gibbons and G.~T.~Horowitz,
Phys.\ Rev.\ D {\bf 53}, 7115 (1996)
[arXiv:hep-th/9512154]\\
F.~Dowker, J.~P.~Gauntlett, S.~B.~Giddings and G.~T.~Horowitz,
Phys.\ Rev.\ D {\bf 50}, 2662 (1994)
[arXiv:hep-th/9312172].\\
F.~Dowker, J.~P.~Gauntlett, D.~A.~Kastor and J.~Traschen,
Phys.\ Rev.\ D {\bf 49}, 2909 (1994)
[arXiv:hep-th/9309075].

\bibitem{Garfinkle}
D.~Garfinkle, S.~B.~Giddings and A.~Strominger,
Phys.\ Rev.\ D {\bf 49}, 958 (1994)
[arXiv:gr-qc/9306023].\\
D.~Garfinkle and A.~Strominger,
Phys.\ Lett.\ B {\bf 256}, 146 (1991).

\bibitem{Hawking1}
S.~W.~Hawking and S.~F.~Ross,
Phys.\ Rev.\ Lett.\  {\bf 75}, 3382 (1995)
[arXiv:gr-qc/9506020].\\
S.~W.~Hawking and S.~F.~Ross,
Phys.\ Rev.\ D {\bf 52}, 5865 (1995)
[arXiv:hep-th/9504019].\\
S.~W.~Hawking, G.~T.~Horowitz and S.~F.~Ross,
Phys.\ Rev.\ D {\bf 51}, 4302 (1995)
[arXiv:gr-qc/9409013].

\bibitem{Emparan}
R.~Emparan,
Phys.\ Rev.\ Lett.\  {\bf 75}, 3386 (1995)
[arXiv:gr-qc/9506025].

\bibitem{Eardley}
D.~M.~Eardley, G.~T.~Horowitz, D.~A.~Kastor and J.~Traschen,
Phys.\ Rev.\ Lett.\  {\bf 75}, 3390 (1995)
[arXiv:gr-qc/9506041].

\bibitem{Kiselev}
V.~G.~Kiselev and K.~G.~Selivanov,
JETP Lett.\  {\bf 39}, 85 (1984)
[Pisma Zh.\ Eksp.\ Teor.\ Fiz.\  {\bf 39}, 72 (1984)].

\bibitem{Vilenkin} A.~Vilenkin, E.~P.~S.~Shellard,
Cosmic string and other topological defects, Cambridge University Press, 1994



\end{thebibliography}
\end{document}